\documentclass[
aps,%
12pt,%
final,%
notitlepage,%
oneside,%
onecolumn,%
nobibnotes,%
nofootinbib,%
superscriptaddress,%
noshowpacs,%
centertags,
secnumarabic,]%
{revtex4}

\begin{document}

\title{
Asymptotic freedom and IR freezing in QCD: the role of  gluon paramagnetism}
\author{\firstname{Yu.~A.}~\surname
{Simonov}} \email{simonov@itep.ru}
 \affiliation{Institute of Theoretical and Experimental Physics\\
Moscow,
 Russia}

\newcommand{\beq}{\begin{eqnarray}}
\newcommand{\eeq}{\end{eqnarray}}
\newcommand{\be}{\begin{equation}}
\newcommand{\ee}{\end{equation}}

\def\la{\mathrel{\mathpalette\fun <}}
\def\ga{\mathrel{\mathpalette\fun >}}
\def\fun#1#2{\lower3.6pt\vbox{\baselineskip0pt\lineskip.9pt
\ialign{$\mathsurround=0pt#1\hfil ##\hfil$\crcr#2\crcr\sim\crcr}}}
\newcommand{\veX}{\mbox{\boldmath${\rm X}$}}
\newcommand{{\SD}}{\rm SD}
\newcommand{\pp}{\prime\prime}
\newcommand{\veY}{\mbox{\boldmath${\rm Y}$}}
\newcommand{\vex}{\mbox{\boldmath${\rm x}$}}
\newcommand{\vey}{\mbox{\boldmath${\rm y}$}}
\newcommand{\ver}{\mbox{\boldmath${\rm r}$}}
\newcommand{\vesig}{\mbox{\boldmath${\rm \sigma}$}}
\newcommand{\vedelta}{\mbox{\boldmath${\rm \delta}$}}
\newcommand{\veP}{\mbox{\boldmath${\rm P}$}}
\newcommand{\vep}{\mbox{\boldmath${\rm p}$}}
\newcommand{\veq}{\mbox{\boldmath${\rm q}$}}
\newcommand{\veK}{\mbox{\boldmath${\rm K}$}}
\newcommand{\vez}{\mbox{\boldmath${\rm z}$}}
\newcommand{\veS}{\mbox{\boldmath${\rm S}$}}
\newcommand{\veL}{\mbox{\boldmath${\rm L}$}}
\newcommand{\vem}{\mbox{\boldmath${\rm m}$}}
\newcommand{\veQ}{\mbox{\boldmath${\rm Q}$}}
\newcommand{\vel}{\mbox{\boldmath${\rm l}$}}
\newcommand{\veR}{\mbox{\boldmath${\rm R}$}}
\newcommand{\ves}{\mbox{\boldmath${\rm s}$}}
\newcommand{\vek}{\mbox{\boldmath${\rm k}$}}
\newcommand{\ven}{\mbox{\boldmath${\rm n}$}}
\newcommand{\veu}{\mbox{\boldmath${\rm u}$}}
\newcommand{\vev}{\mbox{\boldmath${\rm v}$}}
\newcommand{\veh}{\mbox{\boldmath${\rm h}$}}
\newcommand{\vew}{\mbox{\boldmath${\rm w}$}}
\newcommand{\verho}{\mbox{\boldmath${\rm \rho}$}}
\newcommand{\vexi}{\mbox{\boldmath${\rm \xi}$}}
\newcommand{\veta}{\mbox{\boldmath${\rm \eta}$}}
\newcommand{\veB}{\mbox{\boldmath${\rm B}$}}
\newcommand{\veH}{\mbox{\boldmath${\rm H}$}}
\newcommand{\veE}{\mbox{\boldmath${\rm E}$}}
\newcommand{\veJ}{\mbox{\boldmath${\rm J}$}}
\newcommand{\veal}{\mbox{\boldmath${\rm \alpha}$}}
\newcommand{\vegam}{\mbox{\boldmath${\rm \gamma}$}}
\newcommand{\vepar}{\mbox{\boldmath${\rm \partial}$}}
\newcommand{\llan}{\langle\langle}
\newcommand{\rran}{\rangle\rangle}
\newcommand{\lan}{\langle}
\newcommand{\ran}{\rangle}

\begin{abstract}

Paramagnetism of gluons is shown to play the basic role  in establishing main
properties of QCD: IR freezing and asymptotic freedom (AF). Starting with
Polyakov background field  approach  the  first terms of background
perturbation theory are calculated and shown to  ensure  not only  the
classical result of AF  but also IR freezing. For the latter only the confining
property of the background is needed, and the effective mass entering the IR
freezing logarithms is calculated in good agreement with phenomenology and
lattice data.

\end{abstract}
\maketitle

\section{Introduction}

\noindent The  notion of  asymptotic Freedom  (AF) is basic in establishing QCD
as a selfconsistent theory \cite{1}. The extrapolation  of  the QCD  coupling
constant $\alpha_s$ $(Q)$ to larger distances  (smaller momenta $Q$) leads
however to inconsistencies  of several kinds in the pure (nonbackground)
perturbation theory:
\begin{enumerate}
    \item  The appearance  of Landau ghost pole (and other singularities in
    higher orders) precludes  extrapolation to small $Q$ \cite{2}.
    \item IR renormalons make the whole perturbation series not summable even
    in the Borel sense \cite{3}.
    \item The treatment of perturbation series in the Minkowski  space-time
    has  difficulties  and should be reformulated \cite{4}.
\end{enumerate}

    At the same time the IR behavior of $\alpha_s$ in experiment \cite{5} and
    on the lattice \cite{6} does not show irregularities in the Euclidean
    region, $Q^2\geq 0$, and is compatible with IR freezing.

 To ensure this nonsingular behavior a special type of theory was  suggested
 \cite{7}, eliminating Landau ghost pole from the beginning, which is
 phenomenologically successful \cite{8}.

 To understand what dynamical mechanism makes QCD perturbation theory
 consistent and  brings in the IR freezing, as seen on lattice and experiment, the Background
 Perturbation Theory (BPTh), formulated earlier in \cite{9}, was considered,
  treating background as a strong collective field with the property
 of  confinement\cite{10}.

 It was shown in \cite{11}, that the basic effect of this confining  background is to make
 $\alpha_s(Q^2)$ finite at  all $Q^2 \geq 0 $ and thus precluding  appearance
 of  Landau ghost pole and IR renormalons. Moreover, the AF logarithm   approximately  keeps its
 form at large $Q^2$, $\ln \frac{Q^2}{\Lambda^2_{\rm QCD}}$, but at small $Q^2$ the argument
 acquires the additional term, which looks like the two-gluon mass $M^2_{2g}$
 yielding $\ln \left|\frac{Q^2+ M^2_{2g}}{\Lambda^2_{\rm QCD}}\right|.$ This type of form was
 suggested before \cite{12, 13}, however in QCD the appearance of gluon mass is
 forbidden by gauge  invariance.  As will be seen  the term $M_{2g}$ actually has the meaning of the  two-gluon  mass, where
  gluons are connected by the  adjoint string. In \cite{11} the IR freezing was considered in
 the  framework of the static $Q\bar Q$ potential, and the exact general form
 of IR behavior  and exact value of $M_{2g}$ were not actually given. In the
 present paper we present  a more general derivation of the IR freezing based
 on the Polyakov background approach \cite{14}, where the basic one-loop
 element is the scalar self-energy (gluon loop) operator $\Pi(Q^2)$.  As will
 be shown, in the  confining background $\Pi(Q^2)$ acquires the two-gluon
  mass $M_{2g} \approx 2 $ GeV, and ensures both AF at large $Q^2$ and IR freezing at small
 $Q^2$. The explicit  value of this mass is estimated and it is shown, that
 numerically IR freezing is not universal: the IR freezing behavior  and mass (to be
 called IR mass) depends on the embedding process. For comparison the
 background
 perturbation theory for  the  static $Q\bar Q$ system is considered in the one-loop
 approximation and it is shown, that the corresponding IR mass is much lower,
 $M_{2g}^{Q\bar Q}  \approx 1 $ GeV. This latter value is in good agreement with
 phenomenological description of IR freezing \cite{15,16} as well as with lattice  determinations of $\alpha_s$ \cite{17,18}.
  The generalization
 can be  considered as well, leading to the inclusion of multigluon states in
 the asymptotics of a corresponding Green's function, which coupled by
 confinement. Thus all theory becomes finite and
 devoid of IR renormalons \cite{11,19}, hence well defined in the Euclidean region.

 To go beyond Euclidean region, one needs to define better the singularity
 structure of the perturbative series. The logarithmic singularities of the
 free PTh are not physical as well as those in BPTh. To simplify  matter one
 can take the limit $N_c\to \infty $,  where  all QCD amplitudes
 contain only poles \cite{20}.

 The corresponding extrapolation was done in \cite{21,22} where it was shown,
 that equidistant  mass squared spectra of hadrons allow to replace all logs by Euler
 $\psi$-functions, and thus obtain for finite $Q^2<0$ simple poles in
 Minkovskian region, while for large $Q^2>0$ in Euclidean region one has standard logarithmic terms.
 The whole scheme works nicely for both $\beta(\alpha_s)$ and $\alpha_s$ and
 agrees well both with lattice and phenomenology \cite{21,22}.

In all these considerations the nonpositive  definiteness  of the
$\beta$-function of $SU(N_c)$ theory is crucial, and the latter is due to gluon
paramagnetic terms in Lagrangian and gluon Green's function.
 The plan of the paper is as follows.
Section 2  is devoted to the extrapolation of the Polyakov method to the IR
region. Section 3 contains similar treatment for the static potential $Q\bar Q$
system.  In section 4 summary and discussion of results is given.

\section{One-loop evolution of $\alpha_s$ by the Polyakov method}

\noindent As  in \cite{14}, one starts with the gluonic action
$S=\frac{1}{4g^2} \int F^a_{\mu\nu} F^a_{\mu\nu}d^4x$, defined at the scale
$R_1$ (momentum scale $\lambda_1=1/R_1$) and consider Wilson transformation to
the scale $R_2(\lambda_2)$,which can be considered as the change of the
effective integral volume in $S$  from $R_1^4$ to  $R^4_2$. Separating gluon
field into valence gluons $a_\mu$  and backgound $B_\mu$. \be A_\mu =a_\mu
+B_\mu,\label{1}\ee one can expand in $a_\mu$, keeping quadratic in $a_\mu$
terms \be F^a_{\mu\nu} F^a_{\mu\nu}= a^a_\mu \left((D^2_\lambda)_{ab} a^b_\nu
-2g F^a_{\mu\nu} (B) a^b_\mu a_\nu^c f^{abc}\right).\label{2}\ee

As Polyakov mentions, the first term, proportional to $D^2_\lambda$, gives rise
to diamagnetic interaction of valence gluon with background, $g a^a_\mu
B^b_\lambda \partial_\lambda a^c_\nu f^{abc}$, while the second term is
paramagnetic interaction of gluon spins with background. Both can be expressed
in second order through the scalar gluon self-energy $\Pi (x-y)$, which
corresponds  to the loop diagram of two massless scalars, and in case of no
background is \be \Pi_0 (x) = G^2_0 (x) = \frac{1}{(2\pi)^4 x^4}.\label{3}\ee
The resulting expression for the change of one-loop  correction from the scale
$R_1$ to the scale $R_2$ is \be \frac{1}{g^2(R_2)} = \frac{1}{g^2(R_1)} +
\frac{\bar b_0}{4\pi} f(R_1, R_2), \bar b_0 = \frac{11}{3} N_c\label{4}\ee
where we have defined \be f(R_1, R_2) = \int^{|x-y|=R_2}_{|x-y|=R_1} d^4
(x-y)\Pi (x-y)\label{5}\ee which yields the standard expression in the  free
case (no background), \be f_0 (R_1, R_2) = -\frac{1}{4\pi}\ln \frac{R_2}{R_1}.
\label{6}\ee It is   this behavior, which produces AF at small $R_i$ and Landau
ghost pole  \cite{2} appears  when the r.h.s. of (\ref{4}) vanishes. In case of
nonzero background it is necessary to take into account, that ``scalar gluon"
propagator in background, $G(x)$ is no nore free and massless. Moreover, if one
takes into account confinement, then the product $G^2 (x)$ should be replaced
by the two-gluon white Green's function, i.e. the two-gluon glueball Green's
function. $G_{2g} (x)$, and the resulting evolution function
$$ f(R_1, R_2) \to f_{2g} (R_1, R_2),$$
\be f_{2g} (R_1, R_2)\sim \int^{|x|=R_2}_{|x|=R_1} d^4 x \Pi_{2g}
(x).\label{7}\ee

It is our purpose below in this section to calculate $f_{2g}$ both in
coordinate and in the momentum space, proving the  IR freezing in an explicit
way.

To proceed we shall use the exact Fock-Feynman-Schwinger Representation (FFSR)
\cite{23}, for the $2g$ Green's function $\Pi (x,y)$ in the nonzero background
\be \Pi(x,y) = \int^\infty_0 ds_1 \int^\infty_0 ds_2 (Dz^{(1)})_{xy}
(Dz^{(2)})_{xy} e^{-K_1-K_2} W_\sigma (x,y)\label{8}\ee where $K_i =\frac14
\int^{s_i}_0 \left(\frac{dz^{(i)}_\mu}{dt}\right)^2 d\tau_i,$ and $W_\sigma$ is
the Wilson loop with paramagnetic gluon spin insertions, \be W_\sigma (x,y) =
P\exp \left(ig \int_{C(x,y)} A_\mu dz_\mu\right) \exp \left(2 ig \int^s_0 F (z
(\tau))\right) d\tau.\label{9}\ee Here $C(x,y)$ is the loop contour  formed by
the    paths of two gluons from the point $x$ to the point $y$. Averaging over
the vacuum configurations one obtains $\bar \Pi (x,y)$, which is expressed only
in terms of einbein parameters  to be found from the $2 g$ Hamiltonian
\cite{24, 25}, (see Appendix 3 of \cite{11} for details) \be \bar\Pi (x,y) =
\frac{1}{4(2\pi)^{5/2}} \int^\infty_0 \int^\infty_0\frac{d\mu_1 d\mu_2
e^{-\frac{\mu_1+\mu_2}{2} T}}{\tilde \mu^{3/2}\sqrt{T}} G(0,0, T)\label{10}\ee
where we have defined $T\equiv |x-y|, \tilde\mu=
\frac{\mu_1\mu_2}{\mu_1+\mu_2}$, \be G(0,0,T) = \lan 0 |e^{-H_{2g}T}|0\ran =
\sum^\infty_{n=0}  |\psi_{2g}^{(n)} (0) |^2e^{-E_{2g}^{(n)} T}.\label{11}\ee

Here $H_{2g}$ is the $2g$ Hamiltonian with confinement and spin-dependent
interaction, derived in \cite{24,25}, and $``0"$ refers to zero intergluon
distance in the initial and final state.

The $S$-wave spectrum of the Hamiltonian to the lowest order in spin splittings
is well known \cite{24,25,26} \be M_n^{(\mu_1,\mu_2)} =\frac{\mu_1+\mu_2}{2}
+\varepsilon_n (\tilde\mu), ~~ \varepsilon_n (\tilde \mu) = (2\tilde
\mu)^{-1/3} \sigma_{\rm adj}^{2/3} a(n), ~~ a(n)  \approx
\left(\frac{3\pi}{2}\right)^{2/3} \left( n+\frac12\right)^{2/3}.\label{12}\ee

Here $\sigma_{\rm adj} =\frac94 \sigma_{\rm found}$ is the gluonic string
tension, and it is conceivable, that this  gluonic string does not decay for
$N_c\to \infty$, however even for finite $N_c$ the main results are not
sensitive to the high excitations, as will be seen, and hence to the string
decay.
 Inserting
$E_{2g} (n) =M_n$ from (\ref{12}) and $|\psi^{(n)}_{2g} (0)|^2 =
\frac{\sigma_{adj} \tilde\mu}{4\pi},$ \cite{27}, one obtains \be \bar \Pi(x,y)
= A\sum_n \int \frac{ d\mu_1 d\mu_2}{\sqrt{\tilde\mu T}} e^{-M_n
(\mu_1,\mu_2)T}\label{13}\ee where $A= \frac{\sigma}{4(2\pi)^{7/2}}$.

Following \cite{11}, for large $T$ one can  do integration over $d\mu_1,d\mu_2$
using the steepest descent method, which yields stationary point
$\mu_1^{(0)}=\mu_2^{(0)} =\frac14 \bar M_n$, where \be \bar M_n = 4
\sqrt{\sigma_{\rm adj}} \left(\frac{a(n)}{3}\right)^{3/4}\label{14}\ee and the
resulting form of $\Pi(x,y)$ is \be \bar \Pi_{IR} (x,y)
=\frac{A\pi\sqrt{3}}{T^{3/2}}\sum_n \sqrt{\bar M_n} e^{-\bar
M_nT}.\label{15}\ee At small $T$ one instead goes from the sum over $n$ to the
integral, which yields \be \bar \Pi_{AF} (x,y) =\frac{1}{4(2\pi)^4 T^2}
\int^\infty_0 d\mu_1\int^\infty_0 d\mu_2 e^{-\frac{\mu_1+\mu_2}{2} T}=
\frac{1}{16\pi^4 T^4}\label{16}\ee which reproduces the free result (\ref{3}).

Let us now turn  to the momentum space. The analysis of Polyakov in the free
case can be written in the form \be \frac{1}{g^2 (Q^2)} =
\frac{1}{g^2(\mu^2_0)} - \Pi_{\rm para}(Q^2) + \Pi_{\rm dia}
(Q^2),\label{17}\ee where \be \Pi_{\rm para} (Q^2) = 4N_c \Pi (Q^2),~~ \Pi_{\rm
dia} (Q^2) =\frac{N_c}{3} \Pi (Q^2),\label{18}\ee and $\Pi(Q^2)$ in   the free
case is simply a scalar gluon loop, which after renormalization takes the form
\be \Pi_{\rm free} (Q^2) = \int_{\mu_0} \frac{d^4p}{(2\pi)^4 p^2 (p+Q)^2}
=-\frac{1}{16\pi^2} \ln \left(\frac{Q^2}{\mu^2_0}\right).\label{19}\ee

Let us now turn to the case of perturbation theory in the confining vacuum. In
this case two gluons in the loop form bound states, and  we can use the
spectrum, given in (\ref{14}) for large $n$ \be \bar M_n =
\frac{8\pi\sigma_a}{\sqrt{3}} \left( n+\frac12\right) = m^2 +cn, ~~
c=4\pi\sigma_a\left(\frac{2}{\sqrt{3}}\right), ~~ m^2 =
\frac{4\pi\sigma_a}{\sqrt{3}}.\label{20}\ee

For the WKB spectrum in the linear  potential  $\sigma_ar$ one would obtain
instead \cite{28} \be m^2_{\rm WKB} = 2 \pi \sigma_a, ~~ c_{\rm WKB} = 4\pi
\sigma_a\label{21}\ee and we shall exploit these values (differing by 15\% from
$m^2$ and $c$ respectively) in what follows.

In  terms of the bound glueball states $\Pi(Q^2) \to \Pi_{\rm conf} (Q^2)$  can
be written as \cite{29} \be \Pi_{\rm conf} (Q^2) = \sum^\infty_{n=0}
\frac{f_n^2}{Q^2 + M^2_n}<~~ f^2_n = \frac{M_n}{4\mu^2} |\psi^{(n)}_{2g} (0)
|^2=\frac{\sigma_a}{4\pi}\label{22}\ee Replacing in (\ref{22}) the sum over $n$
by the integral and renormalizing  the integral in the same way as in
(\ref{19}) one  obtains \be \Pi_{\rm conf} (Q^2,\mu^2)=-\frac{1}{16\pi^2}\ln
\frac{Q^2+m^2_{\rm WKB}}{\mu^2_0}\label{23}\ee This latter form  coincides at
large $Q^2$ with the perturbation theory result \be \left.\Pi_{\rm  conf}
(Q^2,\mu^2)\right|_{Q^2\to \infty}=\Pi_{\rm free} (Q^2) = -\frac{1}{16\pi^2}
\ln \left( \frac{Q^2}{\mu^2}\right) \label{24}\ee Thus the charge  evolution to
the leading order in the confined  vacuum can be written as \be
\frac{1}{g^2(Q^2)}=\frac{1}{g^2(\mu^2)} - \frac{11}{3} N_c \Pi_{\rm conf}
(Q^2,\mu^2).\label{25}\ee

One can see from (\ref{27}), that  for large $\frac{Q^2+m^2}{\mu^2}$, $\Pi_{\rm
conf} (Q^2,\mu^2)< 0$  and the  AF appears, $g^2(Q^2) < g^2(\mu^2)$.

For nonasymptotically large $Q^2$, but still when $\frac{Q^2+m^2}{\mu^2}$ is
large enough, the standard form of one-loop result for $\alpha_s(Q^2)$ using
(\ref{23}), (\ref{25}) acquires the form \be \alpha_s^{\rm
conf}(Q^2)=\frac{4\pi}{b_0\ln \frac{Q^2+m^2}{\Lambda^2}}.\label{26}\ee

This form  coincides with the one, proposed long ago in \cite{12,13}, where $m$
was associated with the effective mass of two gluons, As one can see, this
notion of mass is  in reality extended to the ground  state mass  of two gluons
connected by the adjoint string, i.e. a  ground-state glueball mass
$M_{2g}(0^{++}).$

Numerically, however, $M_{2g}(0^{++})$ is large, from (\ref{21})
$M_{2g}(0^{++})= m_{\rm WKB} = 1.6$ GeV, which agrees with explicit
calculations in \cite{24,25}, while phenomenological estimate for $m$ in the
$\alpha_s$, entering the static $Q\bar Q$ potential, is $m\approx 1 $
 GeV\cite{15,16}. In the next section we consider this situation in detail and shall find
 $m_{Q\bar Q}$ for the static potential.

 \section{One-loop evolution of $\alpha_s$ for the  static $Q\bar Q$ potential}

 \noindent The purely perturbative derivation of static potential is given in  \cite{30}; for the case of
 confinement this situation was considered in detail in \cite{11}.
 Below we shall give the main results of \cite{11} and, as a new element, we
 estimate numerically the IR freezing mass $m$ in Eq. (\ref{26}) for the
 explicit case of the static  $Q\bar Q$ potential, to be called $m_{Q\bar Q}$.

 One starts with the Wilson loop with rectangular contour $C$ of size $R\times
 T$ containing both nonperturbative confining  background $B_\mu$ and valence
 gluons $a_\mu$. Expanding in powers of $(ga_\mu)$, one obtains a series of diagrams
 with valence gluon exchanges in the background field $B_\mu$ and after vacuum
 averaging
 \be \lan W(B+a)\ran_{B,a} = \exp (- V(R) T - {\rm perimeter})\label{27}\ee   one
 has
 terms
 \be V(R) = V_q (R) + \alpha_s V_2 (R) + \alpha^2_s V_4 (R)+...~.\label{28}\ee
Here $V_q (R)=\sigma_f R$ at large $R$, while $V_n (R), n\geq 2$, contains up
 to $n$ gluons propagating inside the minimal surface $S$ bounded by the
 contour $C$.

 Thus $V_2(R)$ corresponds to the one-gluon exchange; while $V_4(R)$ contains a
 gluon loop (minus ghost loop) on the gluon propagator, a triangle vertex part
 and double gluon exchange  and we take the limit $N_c\to \infty$, so that each gluon line is represented as a double
 fundamental  line. We also take the limit $T\to \infty$ to define static
 potential properly. In this case the change in the  area of the minimal
 surface $S$ due to gluon propagation is only due to inner closed loops in $V_4(R)$, while
 $V_2(R)$ and all single gluon lines are unaffected by confinement,
 \be V_2 (R) = - g^2\frac{C_2 (f)}{4\pi^2} \int^T_0 \int^T_0 \frac{dx_4
 dy_4}{(x_4 -y_4)^2+R^2}=-\frac{\alpha_s^{(0)} C_2(f)T}{R}.\label{29}\ee
 We can write the contribution of $V_2 +V_4 $ in the form
 \be V_2 (R) +V_4 (R) = - \frac{C_2 (f) \alpha_s^{(0)}}{R}
 (1+\alpha_s^{(0)}f(R)),\label{30}\ee
 where $f(R)$ was computed in case of no confinement in \cite{11,30}
 \be f_0(R) =\frac{b_0}{4\pi} \ln \left(\frac{R}{\delta}\right)^2,~~
 \delta\sim 1/\mu.\label{31}\ee

 In the $\overline{MS}$ scheme $ f_0(R)$ was found to be \cite{31}
 \be f_0^{ \overline{MS}}  (R)= \frac{b_0}{4\pi} \left(\ln 9\mu^2R^2)+
 2\gamma_E\right) + \frac{1}{\pi} \left( \frac{5}{12} b_0 -\frac23
 N_c\right).\label{32}\ee
 When confinement is included in the background, $f_{\rm conf} (R)$ is
 expressed through the gluon selfenergy term $\bar \Pi (x-y)$, introduced in
 the previous section.
 \be v_{\rm conf} (R) \equiv \frac{f_{\rm conf} (R)}{R} = \frac{\bar
 b_0}{4\pi^2} \int \frac{d^4r \bar \Pi(r)}{|\veR-\ver|},~~ \bar b_0 = \frac{11}{3}
 N_c.\label{33}\ee
 In the momentum  space the Fourier transform of the $f_{\rm conf} /R$ can be
 written as
 \be \tilde v_{\rm conf} (\veQ) = \int d^3 R v_{\rm conf}(R) e^{i\veQ\veR}
 =\frac{\bar b_0}{\pi} \frac{\Pi(\veQ)}{\veQ^2},\label{34}\ee
 and using (\ref{23}) this can be written as
 \be \tilde v_{\rm conf} (\veQ) =-\frac{\bar b_0}{16\pi^3\veQ^2} \ln
 \frac{\veQ^2+m^2}{\mu^2}\label{35}\ee
Hence the total one-loop potential in momentum space has the form \be \tilde
V_2 (\veQ) +\tilde V_4(\veQ) =-\frac{C_2(f) \alpha^{(0)}_s}{4\pi^2\veQ^2}
\left( 1- \frac{\bar b_0}{4\pi} \alpha_s \ln \frac{\veQ^2+m^2_{Q\bar
Q}}{\mu^2}\right).\label{36}\ee It is now essential, that the string,
connecting the gluons in the internal loop, is fundamental, and therefore \be
m^2_{Q\bar Q} = 2\pi \sigma_f, ~~ m_{Q\bar Q} = 1.06 {\rm ~ GeV}.\label{37}\ee

This is important result, since the IR freezing of the gluon-exchange potential
is essential for the quark model calculations, e.g. of hadron masses (see
\cite{15,16,32}), as well as  in different QCD processes; it can also be tested
on the lattice. In the next section we shall compare the result of (\ref{37})
with other approaches.

\section{Results and discussion}

\noindent  We have studied in previous sections the $\alpha_s$ renormalization
with and without confinement in two different settings: the Polyakov background
setting in section 2 and  the $Q\bar Q$ interaction in section 3. We have
found, that in the quenched ($N_c\to \infty)$ case the one-loop $\alpha_s$ is
given by the same equation in all cases \be \alpha_s (Q^2) = \frac{4\pi}{\bar
b_0 \ln \frac{Q^2+m^2}{\Lambda^2}}, ~~ \bar b_0 =\frac{11}{3} N_c.\label{38}\ee
Here $ m^2=0$ for the case of no confinement in both types of setting,
confirming that the AF is a universal phenomenon. However, the IR mass $m$  is
not universal, it is $m^2\equiv m^2_{gg} = 2\pi\sigma_a$ in case of Polyakov
background approach and $m^2\equiv m^2_{Q\bar Q}= 2\pi\sigma_f$ in case of the
$Q\bar Q$ potential, and the latter can be written as \be \tilde V_{Q\bar Q}
(Q) = - 4 \pi C_2 (f) \frac{\alpha_V (Q)}{Q^2},\label{39}\ee and $\alpha_V (Q)$
to one loop is the same, as in (\ref{38}) with $\Lambda\to \Lambda_V$, while in
the two-loop approximation can be written as \cite{17,18} \be \alpha^{(2)}_V
(Q) = \frac{4\pi}{\bar b_0 t_B} \left( 1- \frac{\bar b_1}{\bar b_0^2} \frac{\ln
t_B}{t_B}\right)\label{40}\ee with $t_B \equiv \ln \frac{Q^2+m^2}{\Lambda_V},
~~ \bar b_1 =102, ~~ \bar b_0 =11.$

The coordinate-space representation $V_{Q\bar Q}(r)$ was studied in detail in
\cite{18} and it was shown, that to a reasonable accuracy (better than 10\% for
$r\geq 0.2$ fm) $\tilde \alpha_V (r)$ can  be approximated by $\alpha_V
(Q=1/r),$ so that   \be V_{Q\bar Q} (r) =- \frac{C_2(f)}{r} \frac{4\pi}{\bar
b_0\ln \left( \frac{1/r^2+ m^2_{Q\bar Q}}{\Lambda^2_V}\right)}\equiv-
\frac{C_2(f)}{r} \tilde\alpha^{(1)}_V(r).\label{41}\ee

In  the two-loop case $V_{Q\bar Q} (r)$ was computed in \cite{17,18}  and is
given by the same Eq.(\ref{41}) where now the two-loop $\tilde\alpha^{(2)}_V
(r)$ is \cite{18} \be \tilde \alpha_V^{(2)} (r) =\tilde \alpha^{(1)}_V (r)
\left\{ 1+ B_1 (r) \frac{\alpha_V^{(1)}(r)}{4\pi}+ B_2 (r)
\left(\frac{\alpha_V^{(1)}(r)}{4\pi}\right)^2\right\},\label{42}\ee where
$B_1(r) =a_1+ 2\gamma_1 (r) \bar b_0,~~B_2(r) =a_2 +2\gamma_1 (r) (\bar
b_1+2\bar b_0 a_1)+\bar b^2_0 \gamma_2(r)$ and $$\gamma_n (r) =\frac{2}{\pi}
\int^\infty_0 dx \frac{\sin x}{x}(\tilde t(x))^n, ~~ \tilde t{(x)} =\ln
\left(\frac{1+m^2r^2}{x^2+m^2r^2}\right);$$
$$\bar a_1 =\frac{31}{3},~~ \bar a_2 = \left[ \frac{4343}{162} +4\pi^2
-\frac{\pi^4}{4} +\frac{22}{3} \zeta(3)\right].$$ One can compare the one-loop
results in $p$ space (\ref{38}), or two-loop result (\ref{40}), and the
corresponding $x$-space expressions (\ref{41}) and (\ref{42}) with lattice and
experiment. Lattice data for $\alpha_V (r) $ from \cite{33} were  compared with
$\tilde \alpha_V^{(2)} (r)$ in \cite{17} and shown to be in good agreement in
the region  0.04 fm $\leq r\leq$ 0.4 fm, measured in \cite{33}, while the
purely perturbative $\alpha^{(2)}_{\rm pert} (r)$  strongly deviates from
lattice data already for $r>0.08$ fm. a similar good agreement can be deduced,
comparing $\alpha_V^{(2)} (r)$ to the Schroedinger functional lattice method
(second reference in \cite{6}). On the phenomenological side, hadron spectra
and especially fine-structure splittings are sensitive to the behavior of
$\alpha_V(r)$. The saturated (frozen) value of $\alpha_V= \alpha_{\rm crit}$ at
large $r$ was assumed in the detailed calculations in \cite{32} and this value
of $\alpha_{\rm crit}$ agrees well with found in \cite{17,18}. Moreover, the
analysis of the splittings between low-lying levels in bottomonium \cite{15},
yields $\alpha_{\rm crit} =0.58\pm 0.02$ in striking agreement with \cite{32}.

A detailed analysis of bottomonium splittings in comparison with lattice data
and experiment \cite{16} proves, that the potential $V_{Q\bar Q} (r)$
(\ref{41}) with $\tilde \alpha_V^{(2)}(r)$ obtained in the confining background
perturbation theory with $m_{Q\bar Q} =1$ GeV is in good agreement  with
experiment. This  confirms the agreement  between the calculations of the
present paper and the physical reality.

Concluding one should stress the crucial role of gluon paramagnetism in
creating  the properties of AF and IR freezing. The  correct  sign of the
logarithmic term in (\ref{26}) is important for its Minkowskian extrapolation
in the form of the sum of pole terms in the $\psi$-function, which in  its turn
correspond to correct physical poles in $\alpha_s$, as shown in \cite{21}.
Moreover,  as shown in \cite{34}, the gluon paramagnetism is responsible  for
the correct physical behavior of the field correlator $\lan F^a_{\mu\nu}(x)
F^a_{\mu\nu}(x)\ran$, implying nonzero and positive value of gluonic
condensate.

 The author is grateful to A.M.Badalian for useful discussions;
financial support of the RFBR grant 09-02-00629a is acknowledged.


\begin{thebibliography}{99}

\bibitem{1} D.~J.~Gross and F.~Wilczek, Phys. Rev. Lett., {\bf 30}, 1323 (1973);
H.~D.~Politzer, Phys. Rev.  Lett.,  {\bf 30}, 1346 (1973); M.~E.~Peskin and
D.~V.~Schroeder, {\it  An Introduction to Quantum Field Theory},
(Addison-Wesley, Reading, USA,
 1995).

\bibitem{2}L.~D.~Landau, A.~A.~Abrikosov and I.~M.~Khalatnikov, Dokl. AN SSSR {\bf 95},
773 (1954).
\bibitem{3}G.'tHooft, in: {\it The Whys of Subnuclear Physics,} Ed. A.~Zichichi, (Plenum Press, New
York, London, 1977);  G.~Parisi, Phys. Lett.  B {\bf 76}, 65 (1977),
B.~Lautrup, Phys. Lett. B {\bf 69}, 105 (1977).

\bibitem{4} D.~V.~Shirkov, Phys. Part. Nucl. Lett. {\bf 5}, 489 (2008); Theor. Math. Phys. {\bf 136}, 893 (2003);
 D.~V.~Shirkov,
hep-ph/0012283.
\bibitem{5} G.~Dissertori and G.~P.~Salam, {\it Quantum Chromodynsmics}, the review
article in:  K.~Nakamura {\it et al.}, JPG {\bf 37}, 075021 (2010); S.~Bethke,
arXiv:0908.1135 [hep-ph].

\bibitem{6}C.~T.~H.~Davies {\it et al.}, (HPQCD Collaboration), Phys. Rev. D  {\bf 78},
114507 (2008); B.~Lucini and G.~Moraitis, Phys. Lett. B {\bf 668}, 226 (2008).

\bibitem{7} D.~V.~Shirkov  and I.~L.~Solovtsov, JINR Rapid Comm. {\bf 2}, 76
(1996); Phys. Rev. Lett. {\bf 79}, 1209 (1997).
\bibitem{8}M.~Baldicchi, A.~V.~Nesterenko, G.~M.~Prosperi,  D.~V.~Shirkov and
C.~Simolo, Phys. Rev. Lett. {\bf 99}, 242001 (2007);  A.~P.~Bakulev,
S.~V.~Mikhailov and N.~G.~Stefanis, JHEP {\bf 1006}, 085 (2010).
\bibitem{9}B.~S.~De Witt, Phys. Rev. {\bf 162}, 1195, 1239 (1967); I.~Honerkamp,
Nucl. Phys. B {\bf 48}, 269 (1972); G.'tHooft, Nucl. Phys. {\bf 62}, 444
(1973); L.~F.~ Abbot, Nucl. Phys. B {\bf 185}, 189 (1981).

\bibitem{10}Yu.~A.~Simonov, in {\it  Lecture Notes in Physics}, (Springer-Verlag,
Berlin, Heidelberg, 1996), Vol. 479, P.139; A.~M.~Badalian and Yu.~A.~Simonov,
Phys. At. Nucl. {\bf 60}, 630 (1997).

\bibitem{11} Yu.~A.~Simonov, Phys. At. Nucl. {\bf 58}, 107 (1995),
hep-ph/9311247.
\bibitem{12}G.~Parisi and R.~Petronzio, Phys. Lett. B {\bf 94}, 51 (1980).

\bibitem{13}J.~M.~Cornwall, Phys. Rev. D {\bf 26}, 1453 (1982); A.~C.~Mattingly
and  P~.M.~Stevenson, Phys. Rev. D {\bf 49}, 437 (1994).
\bibitem{14}A.~M.~Polyakov, {\it  Gauge fields and strings,} (Hardwood Academic, New York 1987).


\bibitem{15}A.~M.~Badalian, B.~L.~G.~Bakker and A.~I.~Veselov, Yad. Fiz. {\bf 67}, 1392
(2004); Phys. Rev. D {\bf 70}, 016007 (2004), A.~M.~Badalian and
V.~L.~Morgunov, Phys. Rev. D {\bf 60} 116008 (1999).


\bibitem{16}A.~M.~Badalian and  A.~I.~Veselov, Phys. At. Nucl. {\bf 68}, 582 (2005); hep-ph/0302072;
A.~M.~Badalian and  B.~L.~G.~Bakker, Phys. Rev. D {\bf 62}094031 (2000).

\bibitem{17}A.~M.~Badalian and D.~S.~Kuzmenko, Phys. Rev. D {\bf 65}, 016004 (2002).



 \bibitem{18}A.~M.~Badalian,  Phys. At. Nucl. {\bf 63}, 2173 (2000).




\bibitem{19} Yu.~A.~Simonov, Pis'ma Zh. Eksp. Theor. Fiz. {\bf 57}, 513 (1993).
\bibitem{20}G.'tHooft, Nucl. Phys. B {\bf 72}, 461 (1974).
\bibitem{21}Yu.~A.~Simonov, Phys. At. Nucl. {\bf 65}, 135 (2002);
hep-ph/0109081; J.~Nonlin. Math. Phys. {\bf 12}, S 625 (2005).

\bibitem{22}Yu.~A.~Simonov, Phys. At. Nucl. {\bf 66}, 764(2003);
hep-ph/0109159.

\bibitem{23}Yu.~A.~Simonov and   J.~A.~Tjon, Ann. Phys.  {\bf 300}, 54 (2002),
hep-ph/0205165.


\bibitem{24}A.~B.~Kaidalov and  Yu.~A.~Simonov, Phys. Lett  B {\bf 477}, 163 (2000);
hep-ph/9912434; Phys. At. Nucl. {\bf 63}, 1428 (2000); hep-ph/9911291.


\bibitem{25} A.~B.~Kaidalov and  Yu.~A.~Simonov, Phys. Lett  B {\bf 636}, 101 (2006);
hep-ph/0512151.

\bibitem{26}  Yu.~A.~Simonov,  in: {\it QCD: Perturbative or Nonperturbative}; L.~S.~Ferreira, P.~Nogueira and J.~I.~Silva--Marcos eds.,
(World Scientific, Singapore 2001).


\bibitem{27} W.~Lucha, F.~F.~Schoeberl and D.~Gromes, Phys. Rep. {\bf 200}, 127 (1991).


\bibitem{28} Yu.~S.~Kalashnikova, A.~V.~Nefediev and  Yu.~A.~Simonov, Phys. Rev. D {\bf
69}, 014037 (2001), hep-ph/0103274.




\bibitem{29}A.~M.~Badalian, B.~L.~B.~Bakker and  Yu.~A.~Simonov,  Phys. Rev. D {\bf 75},
116001 (2007), hep-ph/0702157.



\bibitem{30}T.~D.~Lee, {\it Particle physics and Introduction to Field Theory,} (Harwood, Chur,
London, New York,  1981), P. 447


\bibitem{31}S.~N.~Gupta, S.~F.~Redford and  W.~W.~Repko, Phys. Rev. D  {\bf 24}, 2309
(1981); ibid. D {\bf 26}, 3305.

\bibitem{32}S.~Godfrey and  N.~Isgur, Phys. Rev. D {\bf 32}, 189 (1985).



\bibitem{33}G.~S.~Bali, Phys. Lett. B {\bf 460}, 170 (1999), hep-ph/9905387.


\bibitem{34} Yu.~A.~Simonov and  V.~I.~Shevchenko, Adv. High En. Physics, 2009; 873061 (2009); arXiv; 0902.1405 [hep-ph];
Yu.~A.~Simonov, Phys. At. Nucl. {\bf 69}, 528 (2006); Yu.~A.~Simonov,
arXiv:1003.3608 [hep-ph].

\end{thebibliography}
 \end{document}